\documentclass{aa}  

\usepackage{amssymb}
\usepackage{aas_macros}
\usepackage{graphicx}
\usepackage{txfonts}

\newcommand{\header}[1]{\multicolumn{1}{c}{\textrm{#1}}}
\newcommand{\Teff}{$T_{eff}$}

\newcommand{\Mo}{$M_{\odot}$}
\newcommand{\Ro}{$R_{\odot}$}
\newcommand{\kms}{km\,s$^{-1}$}
\newcommand{\pz}{PZ~Mon}
\newcommand{\vr}{$v_{\rm rad}$}

\newcommand{\rs}{RS~CVn}
\newcommand{\vs}{$v\sin{i}$}
\newcommand{\ii}{\,{\sc ii}}

\begin{document}
 
\title{PZ Mon is a new synchronous binary with low mass ratio
\thanks{Based on observations collected at the BTA telescope (Special
Astrophysical Observatory, Russia) and Zeiss-2000 telescope (Terskol
observatory of Institute of Astronomy, Russia)}}

\author{Yu.~V.~Pakhomov \inst{1}
\and
N.~A.~Gorynya \inst{1,2}
}

\institute{Institute of Astronomy, Russian Academy of Sciences, Pyatnitskaya 48,
119017, Moscow, Russia\\
\email{pakhomov@inasan.ru}
\and
Lomonosov Moscow State University, Sternberg Astronomical Institute,
Universitetskij prospekt, 13, Moscow 119991, Russia\\
\email{gorynya@sai.msu.ru}
}

\titlerunning{\pz\ is a new synchronous binary with low mass ratio}
\authorrunning{Pakhomov \& Gorynya}

\date{Received September 15, 1996; accepted March 16, 1997}

\abstract
{Analysis of new radial velocity measurements of the active giant \pz\ is
presented.}
{Only in 2015 was reported that \pz\ may be classified as \rs\ giant. At the
same time was discovered the variability of radial velocity. However, lack of
the data is not allowed to determine parameters of the system.}
{The measurements of radial velocity were performed using Radial Velocity Meter
installed at the Simeiz 1-m telescope of the Crimean Astrophysical Observatory
and using echelle spectrographs installed at the 2-m Zeiss telescope of the
Terskol Observatory and the 6-m telescope BTA of the Special Astrophysical
Observatory of the Russian Academy of Sciences.}
{We estimated parameters of this binary system
including the
$\gamma$-velocity 25.5$\pm$0.3~\kms, the period on the circular orbit
$P=34.15\pm0.02$~days, the mass of the secondary component $M_2$=0.14~\Mo, and
the mass ratio $q=0.09$}
{The mass ratio is a smallest value among known \rs\
type giants. Combined with photometric data we conclude that \pz\ is a system
with synchronous rotation, and there is a big cool spotted area on the stellar
surface towards to the secondary component that provides the optical
variability.}

\keywords{
(Stars:) binaries: general --
Stars: individual: PZ Mon --
Stars: kinematics and dynamics --
Stars: variables: general --
(Stars:) starspots
}

\maketitle

\section{Introduction}

\object{PZ Mon} (HD\,289114, $V\approx9$~mag) is active K2III star type of \rs\
located at a distance of about 250~pc \citep{2015MNRAS.446...56P}. Early the
star was classified as a red dwarf with rapid irregular variability
\citep{2009yCat....102025S}, SIMBAD calls this star as a flare star based on
several articles \citep[e.g.][]{1989A&A...217..187P,  1999A&AS..139..555G}.
Nevertheless, the photometric period of about 34~days was detected by
periodogram analysis \citep{2007OAP....20...14B} which was attributed to the
rotational modulation of the spotty star. This assumption was confirmed by
\citet{2015MNRAS.446...56P} through comprehensive analysis of \pz; the measured
rotational velocity 10.5~\kms\ corresponds to the found period 34.14 days and
the inclination of the rotation axis sin~$i$=0.92. Classification of \pz\ as a
\rs\ type star was carried out by analysis of its spectrum which shows evidences
of the chromospheric activity: H$\alpha$ emission, D$_3$ helium absorption,
emission in core of strong lines of sodium and magnesium. The luminosities in UV
and X-ray also agree with the luminosities of other \rs\ stars. The duality of
\pz, main characteristic of \rs\ stars, was discover by observations of
variability of radial velocity with a possible period of about 17~days. However,
absent of the measurements of radial velocity throughout all phases not allowed
to make the final conclusion about true values of the period and the amplitude.
In this work we close this gap by new radial velocity measurements
(Section~\ref{sec-observations}), also presented determination of parameters of
the \pz\ system (Section~\ref{sec-parameters}), analysis these data together
with photometric data (Section~\ref{sec-photometry}), and discussion in
Section~\ref{sec-discussion}.

\section{Observations}
\label{sec-observations}

 The main observations (set \#1) we obtained 2014 Oct.~24--Nov.~9 using Radial
Velocity Meter (RVM) \citep{1987SvA....31...98T} installed at the Simeiz 1-m
telescope of the Crimean Astrophysical Observatory. Zero point velocity was
determined by observations of several IAU velocity standards each night. 

Seven spectra of \pz\ were obtained within two observational sets using the
MAESTRO echelle spectrograph (the resolving power $R$=40\,000) installed in the
coude focus at the 2-m Zeiss telescope of the Terskol Observatory of the
Institute of Astronomy of the Russian Academy of Sciences. The first set (\#2)
contains one spectrum taken at 2014 Dec.~15, the second (\#3) has six spectra
taken at 2015 Jan.~18--26 on the Wright Instruments CCD (1242x1152) (see the
journal of observations in Table~\ref{tab:obs}). The exposure time was from 0.5
up to 2.75~hours (depending on the weather conditions) which was divided into
several parts to reduce influence of cosmic rays. The signal-to-noise ratio
for each spectrum are presented in the last column of Table~\ref{tab:obs}. For
three nights (Dec.~15 and Oct.~23 27) spectrum of the IAU radial velocity
standard $\beta$~Gem was taken after \pz. In additional, for each set we took
spectra of scattered sunlight. Data have been reduced using the MIDAS package
\textit{echelle}. 88 echelle orders were extracted in the region of 3530 --
10060~\AA\ with the wavelengths calibrated via ThAr hallow-cathode lamp. The
spectra normalization was performed with the blaze function obtained from
exposures of the star $\eta$~Uma. Each spectrum of \pz\ was reduced separately,
and then for each night these spectra were combined together with an allowance
for the Earth's rotation. 

One spectrum of \pz\ (set \#4) was obtained sequentially in three exposures at
2015 Feb.~10 using the NES echelle spectrograph with a slicer (the resolving
power $R$=60\,000) installed at the 6-m telescope BTA of the Special
Astrophysical Observatory of the Russian Academy of Sciences on the e2v CCD42-90
(4632x2068). 54 echelle orders were extracted in the region of 3890 -- 6980~\AA\
with the wavelengths calibrated via ThAr hallow-cathode lamp. Similarly, each
spectrum of \pz\ was reduced separately, and then the three spectra were
combined together with an allowance for the Earth's rotation. 

Radial velocity (RV) was determined from \pz\ spectrum in the region of
4800--6300~\AA\ by cross correlation with the spectra of $\beta$~Gem and
scattered sunlight broadened to the rotational velocity of \pz. The result are
presented in Table~\ref{tab:Vr} where the first column denotes the order number
of observational sets (\#1 is from RVM, \#2, \#3, and \#4 are from spectral
data), following are Julian day, the radial velocity \vr\ and the errors. Also
we reprocessed preview spectra of \pz\ taken in 2012 \citep{2015MNRAS.446...56P}
and estimated RV by the same method \vr=28.1$\pm$0.4~\kms.

\begin{table}  
\renewcommand{\tabcolsep}{1.2mm}
\caption{Characteristics of the spectral observations of \pz.\label{tab:obs}}
\centering
\begin{tabular}{cccccc}
\hline
\header{Set}&\header{Data}&\header{Time}&\header{Number} &
\header{Total} & \header{Total} \\
            &             &     UTC     &\header{of} &
\header{exposure}         & \header{S/N} \\
            &             &             &\header{exposures} &
\header{time, sec}      &                \\
\hline 
2 &  2014-12-15 & 22:14-00:21 & 4 & 7200 & 45 \\
3 &  2015-01-18 & 20:28-23:23 & 4 & 9900 & 65 \\
  &  2015-01-19 & 20:59-23:42 & 5 & 9000 & 90 \\
  &  2015-01-20 & 23:26-23:56 & 1 & 1800 & 40 \\
  &  2015-01-23 & 20:19-22:08 & 3 & 5400 & 75 \\
  &  2015-01-24 & 20:21-21:23 & 2 & 3600 & 80 \\
  &  2015-01-26 & 19:23-20:24 & 2 & 3600 & 60 \\
4 &  2015-02-10 & 16:47-18:20 & 3 & 5400 & 120\\
\hline
\end{tabular}
\end{table}

\begin{table}  
\caption{Radial velocities of \pz.\label{tab:Vr}}
\centering
\begin{tabular}{llll}
\hline
\header{Set}&\header{JD}&\vr  & $\sigma$\vr  \\
            & 2450000+  &\kms &~\kms    \\
\hline 
1 & 6954.596 &  29.00 & 0.57 \\
  & 6959.609 &  23.95 & 0.20 \\
  & 6960.594 &  23.42 & 0.18 \\
  & 6961.585 &  21.82 & 0.28 \\
  & 6962.657 &  23.78 & 0.34 \\
  & 6964.596 &  20.55 & 0.32 \\
  & 6965.587 &  19.80 & 0.28 \\
  & 6966.577 &  20.03 & 0.24 \\
  & 6967.562 &  18.87 & 0.23 \\
  & 6968.541 &  21.22 & 0.27 \\
  & 6969.560 &  21.09 & 0.22 \\
  & 6970.547 &  21.55 & 0.25 \\
2 & 7007.458 &  24.6  & 0.5  \\
3 & 7041.368 &  21.7  & 0.5 \\
  & 7042.385 &  24.2  & 0.5 \\
  & 7043.487 &  24.9  & 0.7 \\
  & 7046.357 &  27.8  & 0.5 \\
  & 7047.358 &  28.5  & 0.7 \\
  & 7050.318 &  30.9  & 0.5 \\
4 & 7064.210 &  23.91 & 0.16 \\
\hline
\end{tabular}
\end{table}

\section{Determination parameters of radial velocity curve}
\label{sec-parameters}

\begin{figure}
\centering
\resizebox{0.95\hsize}{!}{\includegraphics[clip]{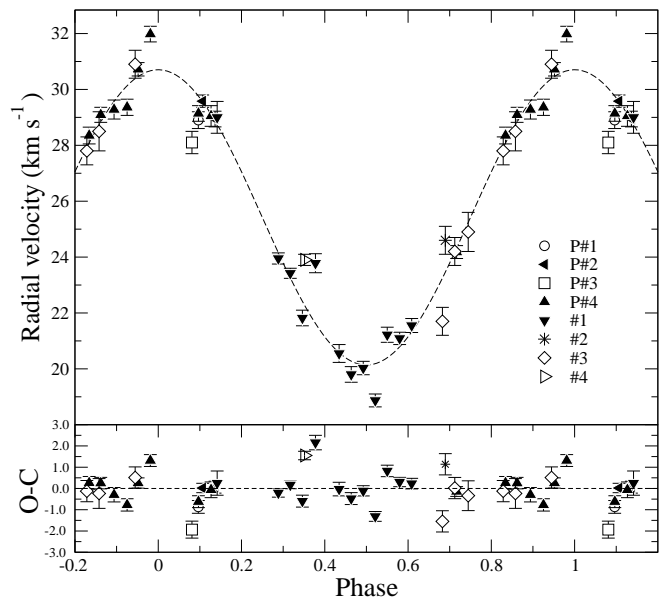}}
\caption{\textit{Top:} Radial velocity measurements convolved with the 
period P=34.15~d. Different sets are marked by different symbols. The labels
correspond to the set number of Table~3 from 
\citep[][started with "P"]{2015MNRAS.446...56P} and Table~\ref{tab:Vr} of
current article. Dashed line is the theoretical radial velocity curve calculated
by derived parameters. \textit{Bottom:} Residual radial velocities (O-C).}
\label{fig:Phase}
\end{figure}

\begin{table}  
\caption{Parameters of \pz\ system. Supposed values marked by a colon.
\label{tab:par}}
\centering
\begin{tabular}{llll}
\hline
\multicolumn{2}{c}{Parameter}& \pz\ A         & \pz\ B \\
\hline
$\gamma$& (\kms) & \multicolumn{2}{c}{25.5$\pm$0.3} \\
$P$   & (\kms)     & \multicolumn{2}{c}{34.15$\pm$0.02} \\
$K$   & (\kms)     & 5.4$\pm$0.4    & 0.5: \\
$\sigma_V$ & (\kms)& 0.9            &      \\
\vs   & (\kms)     & \multicolumn{2}{c}{0.92$\pm$0.25} \\
$M$   & (\Mo)      & 1.5$\pm$0.5    & 0.14$\pm$0.05 \\
$q=M_2/M_1$ &      & \multicolumn{2}{c}{0.09$\pm$0.03} \\  
$R$   & (\Ro)      & 7.7$\pm$1.9    & 0.22: \\
$T$   & (K)        & 4700$\pm$100   & 2700: \\
SpType&            & K2III          & M7V:  \\
$e$   &            &                & 0.0$\pm$0.05 \\
$a$   & (a.u.)     & 0.018$\pm$0.005& 0.24$\pm$0.03 \\
\hline
\end{tabular}
\end{table}

In \citep{2015MNRAS.446...56P} we have described RV variation up to its maximum
value. The set~\#3 shows very similar RV behavior that gives us the possibility
to consider the sets in the common range of phases with the time difference
444.3$\pm$0.5~days. In this way, we probed values of the period as 444.3/$n$
where $n$ is the integer number of orbital turns. For $n=13$ we obtained the
best solution $P=34.18\pm0.04$~days. The period of about 17~days
\citep{2015MNRAS.446...56P} is not consistent with the observational set~\#1,
probably, an error at the key point much more due to low resolution. To adjust
the period value we use the fact that two pairs from the sets \#2 and \#4 from
\citep{2015MNRAS.446...56P} and one point from \#1 of the current work has close
RV and phase values. The estimated $P$=34.15$\pm$0.02~days also agrees with the
earliest RV measurement of \cite{1998IBVS.4580....1S}. All available at this
moment the RV values are combined in Fig.~\ref{fig:Phase}. 

We applied Levenberg-Marquardt method to estimate the parameters of the RV
curve: $\gamma$ is the constant radial velocity of the system \pz\
($\gamma$-velocity), $K$ is the semiamplitude, $e$ is the orbital eccentricity,
$\omega$ is the longitude of periastron. The following formula was used to fit
the observations:
$$
 v_{rad} = \gamma + K \left[ cos (\theta + \omega) + e~cos~\omega \right]
$$
where $\theta$ is the orbital true anomaly. The derived parameters are presented
in Table~\ref{tab:par} together with updated values of physical characteristics
of \pz\ binary system. Note, that the major semiaxis was estimated as
$a_1=KP/(2\pi\,\textrm{sin}~i)$ for the primary component and
$a_2^3=G(M_1+M_2)P^2/(4\pi^2)$ for the secondary component, where we assumed
$i$=$i_{rot}$=$i_{orb}$. The eccentricity $e$=0.0$\pm$0.05, we adopt $e$=0, so
$\omega$ was excluded. The fit curve drawn in Fig.~\ref{fig:Phase} where in
bottom O--C values are presented. The O--C values more than typical error of the
measurements. The average deviation is about of 0.9~\kms, up to 2.0~\kms\ for
two measurements.

The derived ephemeris of RV maximum is
$$
JD = 2457052.2{\scriptstyle\pm0.3} + 34.15{\scriptstyle\pm0.02}~E
$$
 
\section{Photometric data}
\label{sec-photometry}

Since the photometric period of \pz\ is stable and related to the stellar
rotation \citep{2015MNRAS.446...56P} and approximately equal to the RV period we
decided to find a potential relationship between axial rotation of \pz\ and
orbital rotation of the secondary component. 

\begin{figure}
\centering
\resizebox{0.95\hsize}{!}{\includegraphics[clip]{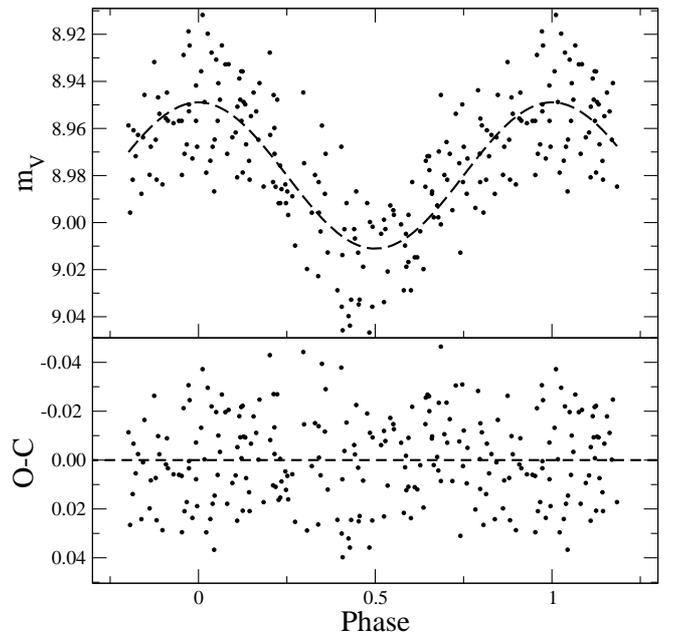}}
\caption{\textit{Top:} Light curve in V band from the ASAS data (2005-2009)
convolved with P=34.13~days and fitted by cosines low (\textit{dashed line}).
\textit{Bottom:} Difference between observations and calculations
($\sigma=0.02^m$).}
\label{fig:PhV}
\end{figure}

\begin{figure}
\centering
\resizebox{0.95\hsize}{!}{\includegraphics[clip]{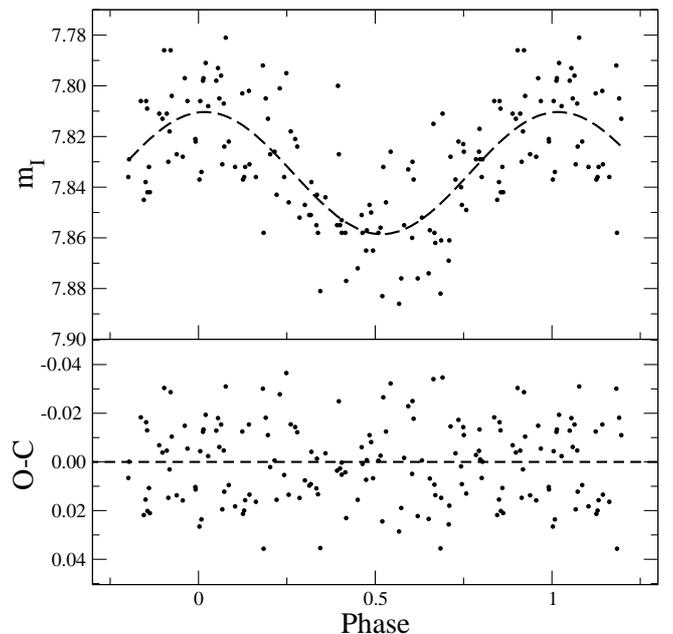}}
\caption{\textit{Top:} Light curve in I band from the ASAS data (2005-2009)
convolved with P=34.13~days and fitted by cosines low (\textit{dashed line}).
\textit{Bottom:} Difference between observations and calculations
($\sigma=0.02^m$).}
\label{fig:PhI}
\end{figure}

To derive the photometric ephemeris we used observations in V and I bands of the
ASAS project \citep[][http://www.astrouw.edu.pl/asas/]{1997AcA....47..467P}. The
period of 2005-2009 (5 sets of observations) is more preferable due to expressed
changes of the magnitude. The light curve convolved with P=34.13~days is
presented in Fig.~\ref{fig:PhV} for V Johnson band and in Fig.~\ref{fig:PhI} for
I band. The typical error of magnitudes given by the ASAS is about of
0.03--0.05$^m$ but really the accuracy may be better that can be seen from the
O--C values. In both case, to avoid the global trend we reduced the magnitudes
to one of the observational sets. The changes in Fig.~\ref{fig:PhV},
\ref{fig:PhI} like the law of cosines with semiamplitude of about 0.03$^m$. The
ephemeris of maximum of brightness is
$$
JD=2454807.2{\scriptstyle\pm0.7} + 34.13{\scriptstyle\pm0.02}~E
$$

\begin{figure}
\centering
\resizebox{0.69\hsize}{!}{\includegraphics[clip]{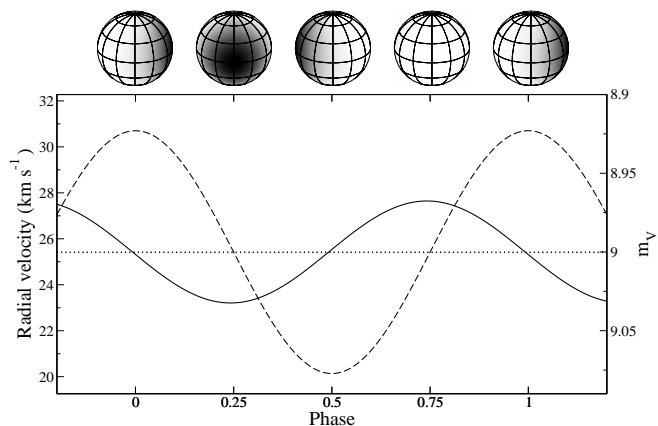}}
\resizebox{0.95\hsize}{!}{\includegraphics[clip]{Shift.eps}}
\caption{Radial velocities (\textit{dashed line}) and photometric
(\textit{thin line}) ephemerides. The last is in assuming a constant of the
average magnitude (9$^m$) and the semiamplitude (0.03$^m$). \textit{Top:} the
apparent position of the main spotted area.}
\label{fig:Ephemeris}
\end{figure}

In Fig.~\ref{fig:Ephemeris} we present relative locations of the light and the
radial velocity curves calculated from derived ephemerides. Because real changes
of magnitudes are more complex due to irregular nature the light curve in
Fig.~\ref{fig:Ephemeris} synthesized with the constant average magnitude of
9$^m$ and the constant semiamplitude of 0.03$^m$. Fig.~\ref{fig:Ephemeris} shows
the phase shift between the curves $\Delta\phi$=0.78$\pm$0.04 for $P$=34.13~days
and 0.74$\pm$0.04 for $P$=34.15~days. Thus, the shift provides the lead of the
photometric curve by one fourth of the period. However, this fact refers to the
average picture. Probably, the deviations in Fig.~\ref{fig:PhV} and
Fig.~\ref{fig:PhI} may partially related to the differential rotation of \pz\
and the change positions of the active regions which are responsible for the
observable variability.

\subsection{Model of \pz\ system}

Minimum of the brightness in Fig.~\ref{fig:Ephemeris} matches to the
$\gamma$-velocity. Such picture may observed if a cool spot on the stellar
surface lie on a line connecting the stars. Probably, this spot is a effect of
the activity due influence of the secondary component and is a main reason of
photometric variability (see top of Fig.~\ref{fig:Ephemeris}). 

Typically, light curve of \rs\ stars shows a flat plateau but, in case of \pz,
it like the law of cosines with semiamplitude of about 0.03$^m$ that may be
caused by big spotted area with size of about semisphere and with the average
photosphere temperature of 40--50~K below relative to opposite stellar surface.

We modelled two simple cases for equatorial spotted area. In the first case, the
spotted area of radius 90 degree has constant intensity $I_{spot}$ relative
unspotted area. In the second case, the spotted area has intensity distributed
by gaussian $I=I_{spot}e^{-(r/90)^2}$, where $r$ is the angular distance from
the spot center. The result of calculations is $I_{spot}$=0.943, 0.915 for these
cases, respectively, that provides observed behavior and the amplitude of the
light curve. Also we calculated influence of the spots on the RV measurements
using given distribution of the intensity. The maximal changes $\Delta$\vr=0.14
and 0.10~\kms\ were obtained which lower than the errors in our measurements. 

If the spotted area represented by evenly distributed spots on the apparent
semisphere then we can calculate the increase of the filling factor on the more
spotted area relative to the opposite side. \citet{2006A&AT...25..247A} have
estimated the filling factor $S_1\approx$20\%\ for the semisphere which provides
maximum of the brightness. The ratios of maximal and minimal intensities is
$\frac{I_2}{I_1}=\frac{1-S_2(1-\beta)}{1-S_1(1-\beta)}$,
where $\beta<<1$ is the ratio of intensity in the spots and real unspotted
stellar surface. Using our value $I_{spot}=\frac{I_2}{I_1}$=0.943 we calculated
$S_2$=24\%\ which slightly depends on $\beta$. Thus, the difference of the
filling factor between semispheres with minimal and maximal spots number is
about 4\%.

\section{Discussion}
\label{sec-discussion}

The parameters of the RV curve point to the secondary star of mass 0.14~\Mo\ at
the circular synchronous orbit with the major semiaxis $a$=0.23~a.u. = 35
millions km = 6.5$R_{PZ}$. The star may be a red dwarf with the effective
temperature \Teff$\approx2700$~K, and the radius $R\approx0.22$~\Ro. In the sky
this is a star of magnitude 19$^m$ located at an angular distance of $0.001''$
from \pz\ that making it invisible object from the Earth. We exclude a
possibility that the variability is caused by an eclipse. In system of \pz\ can
occur eclipse if angle of the inclination more than 81$^\circ$ (sin~$i>$0.988).
In this case, the apparent variation of magnitudes will be too small, about
0.0008$^m$. To provide observed amplitude $\sim$1/18 of the stellar disk should
be closed, then radius of the secondary star should be 0.23~$R_{PZ}$ or 1.8~\Ro\
that is not consistent with the mass of the object. 

We observe difference between phases $\Delta\phi=0.75$ of RV and light curves.
The same effect was observed for other some \rs\ stars.
\cite{2011AJ....141..140B} do not notice the effect in V474~Car but available
data allow us to calculate it, we found $\Delta\phi=0.75$.
\cite{2010A&A...521A..36M} analyzed photometric data along orbital modulation
and found maximum of the brightness around 0.25p and 0.75p depending on the spot
activity. \cite{2007MNRAS.382.1133Z} studied \rs\ type star DV~Psc, the
difference 0.75p between photometric and radial velocity phases is shown in
Fig.~3 of its article. Also we can see narrow flux minimum corresponds to the
RV average value. Authors concluded that these effects can be caused by the cool
spot towards to the secondary component. \cite{2002A&A...389..202O} perform the
doppler mapping of short period synchronous \rs-type giant UZ~Lib. All maps
linked to the orbital ephemeris show minimum of brightness at the phases about
of 90 and 270 degrees, i.e. main spots located towards to the secondary
component and on the opposite site. However, in most cases, doppler mapping of
\rs\ stars does not show stable locations of big spots. To clarify the nature of
\pz\ activity we need the spectral monitoring and the doppler mapping of its
surface. 

This effect of activity is expected. Indeed, the secondary component should
affect to main star which demonstrates chomospherical activity.
\cite{1976ApJ...210L..27O} have detected radio emission from \rs\ binary star
HR~1099. \cite{1986PASAu...6..316B} observed the variations of radio from spots
related to the stellar rotation and explained the source of radio by
gyro-synchrotron radiation. \cite{1984ApJ...282L..23L} using VLBI estimated the
size of radio source as comparable diameter of active star.
\cite{2001A&A...373..181T} observed AR~Lac variable of \rs\ type also using
VLBI. The obtained data indicate a source size close to the separation of the
binary components, suggesting the possibility of an emitting region located
between the system components. Note, many \rs\ type stars were contained in the
FIRST catalog of radio sources \citep{1997ApJ...475..479W}. However, not all
\rs\ giants are sources of radio emission. \pz\ is one of them, it was not
marked as a radio source. \cite{1998ASPC..154.1551R} searched the tidal effects
by radio emission at difference rotational phases for two \rs\ type stars but
not found.

\begin{table*}  
\caption{Comparison of the parameters of \rs\ type analogues of
\pz\ by the spectral type, the period, and the mass.\label{tab:analog}}
\centering
\begin{tabular}{lllllll}
\hline
\multicolumn{2}{c}{Parameter} & \pz & V1379~Aql & AZ~Psc &  V4138~Sgr & FG~Cam\\
\hline
$P_{orb}$     & (days) & 34.15 & 20.66 & 47.12 & 13.05 & 33.83\\
$P_{rot}$     & (days) & 34.13 & 25.64 & 91.2  & 59    & 31.95\\
$V$sin$i$     & (\kms) & 10.5 &$\sim$19&$\sim$4&$\sim5$& 12.4 \\
$K$           & (\kms) & 5.4   & 13.1  & 10.35 & 9.89  & 5.28\\
$M_1$         & (\Mo)  & 1.5   &  2.4  & 1.5   & 1.5   & 3.3\\
$M_2$         & (\Mo)  & 0.14  &  0.31 & 0.24  & 0.24  & 0.38\\
$q=M_2/M_1$   &        & 0.09  &  0.13 & 0.16  & 0.16  & 0.12 \\
SpType$_1$    &        & K2III & K0III & K0III & K1III & K2III\\
SpType$_2$    &        & M7V?  & sdB   & dF    & dB    & M0III\\
$\Delta\,m_V$ & (mag)  & 0.03  & 0.30  & 0.33  & ~0.3  & 0.07\\
distance      & (pc)   & 250   & 270   & 170   & 80    & 257 \\ 
source of     &        &UV,X   &IR,UV,X&IR,UV,X&Rad,IR,UV,X & Rad,IR,X\\
log($L_{bol}/L_\odot$)&& 1.4   & 1.7   & 1.6   & 1.1   & 2.1 \\
log($L_X/L_{bol}$) &   & -4.5  & -4.8  & -4.3  & -4.5  & -4.6\\
log($L_{FUV}/L_{bol}$)&& -4.2  & -1.9  & -3.4  & -4.4  & -4.7\\
log($L_{NUV}/L_{bol}$)&& -3.1  & -2.0  &  --   & -2.8  & -3.3\\
IR excess at  & ($\mu$m)&  --   &60,100 & 100   & 100   & 60,100 \\
\hline
\end{tabular}
\end{table*}

\pz\ has small mass ratio of the components $q=M_2/M_1=0.09$ which is the
smallest value among known \rs\ giants. At the same time, the system is
synchronous. We analyzed available data to find analogues of \pz\ by mass ratio
among long-periodical ($P>20$~d) \rs\ giants. In the General Catalog of Variable
Stars \citep{2009yCat....102025S} there are 20 confirmed giants of \rs\ type,
half of them are long-periodical. While among all \rs\ stars this portion is
about of 8\%. In the Catalog of chromospherically active binary stars
\citep{2008MNRAS.389.1722E} there are 164 \rs\ type stars, 64 among them has
period more than 20~days. While amount of giants is 78 and 52, respectively.
Thus, in the family of long-periodical \rs\ stars giants dominate. We found four
analogues with minimal mass of the secondary component: V1379~Aql \citep[data
of][]{1992MNRAS.258...64J}, AZ~Psc \citep[data of][]{2004A&A...424..727P,
2005AJ....129.1669F} and V4138~Sgr \citep[data of][]{2005AJ....129.1669F}. The
comparison of its parameters together with \pz\ presented in
Table~\ref{tab:analog}. All comparable systems are asynchronous and they shown
more noticeable changes of magnitudes exclude last one FG~Cam. They have the
same spectral class and close values of luminosities ratio of X-rays and
bolometric $L_X/L_{bol}$. Unlike \pz\ they are sources of radio and infrared
emission. Three asynchronous stars may have hot components. V1379~Aql has
apparent in UV hot subdwarf of temperature of about 25\,000--30\,000~K. In UV
region V1379~Aql shows only one chromospheric line Mg\ii; AZ~Psc has noisy
spectrum without visible lines; V4138~Sgr, contrariwise, has strong spectral
lines \citep{2000Ap&SS.273..155W} with the luminosities comparable and exceeding
of \pz. UV luminosities calculated from fluxes in terms of GALEX
\citep{2011Ap&SS.335..161B}: FUV $\lambda_0$=1539~\AA,
$\Delta\lambda$=1344--1786~\AA, NUV $\lambda_0$=2316~\AA,
$\Delta\lambda$=1771--2831~\AA. Perhaps, big amplitude of analogues related to
high temperature of the secondary component. If this assumption is true then
little amplitude of \pz\ and FG~Cam would indicate the low temperature of the
secondary component. All comparable stars show excess in infrared starting from
60--100~$\mu$m that may points to circumstellar shell.

\section{Conclusions}

We performed the radial velocity measurements of the active giant \pz\ and
determined the parameters of this binary system. The mass ratio of the
components $q=0.09$ is a smallest value among known \rs\ giants. Within errors
of the derived periods we conclude that the system is synchronous. However, we
need to improve the conclusion by new photometric and radial velocity
measurements.

\begin{acknowledgements}
 This work is supported by the grant "Nonstationary Phenomena in the Universe"
of the Russian Academy of Sciences. We thank the administration of the Simeiz
Section of the Crimean Astrophysical Observatory for allocating observation time
at the 1-m telescope. This work was partial supported by the Russian Foundation
for Basic Research (projects 15-02-06046, 14-02-00472). Calculations of the 
$\gamma$-velocities were supported by Russian Scientific Foundation (project 
14-22-00041).
\end{acknowledgements}

\bibpunct{(}{)}{;}{a}{}{,} 
\bibliographystyle{aa}
\bibliography{paper}

\end{document}